\documentclass[%
 reprint,
 amsmath,amssymb,
 aps,
]{revtex4-2}

\usepackage{graphicx}
\usepackage{dcolumn}
\usepackage{bm}

\begin{document}

\preprint{APS/123-QED}

\title{A Semi-Empirical Descriptor for \\Open Circuit Voltage}

\author{Sourav Baiju$^{*}$}
 \affiliation{%
 Materials Syntesis and Processing (IMD-2), Institute of Energy Materials and Devices \\ Forschungszentrum Juelich GmbH, Germany\\ MESA+ Institute for Nanotechnology\\ University of Twente \\ 7500 AE Enschede, the Netherlands 
}%
\author{Konstantin Köster$^{*}$}
\thanks{$^{*}$These authors contributed equally to this work.}
 \affiliation{%
 Materials Syntesis and Processing (IMD-2), Institute of Energy Materials and Devices \\ Forschungszentrum Juelich GmbH, Germany\\ MESA+ Institute for Nanotechnology\\ University of Twente \\ 7500 AE Enschede, the Netherlands 
}%

\author{Mark Huijben}%
\affiliation{%
 Materials Syntesis and Processing (IMD-2), Institute of Energy Materials and Devices \\ Forschungszentrum Juelich GmbH, Germany\\ MESA+ Institute for Nanotechnology\\ University of Twente \\ 7500 AE Enschede, the Netherlands 
}%
\author{Payam Kaghazhi}%
 \email{p.kaghazchi@fz-juelich.de}
\affiliation{%
 Materials Syntesis and Processing (IMD-2), Institute of Energy Materials and Devices \\ Forschungszentrum Juelich GmbH, Germany\\ MESA+ Institute for Nanotechnology\\ University of Twente \\ 7500 AE Enschede, the Netherlands 
}%

\date{\today}

\begin{abstract}
     Layered transition metal oxides (TMO) are widely used as cathode materials in Na/Li batteries. The open-circuit voltage (OCV), which determines the energy density (together with capacity), is among the key physical and chemical factors influencing the performance of cathodes. The shape of the voltage profile is also influenced by the formation energy of intermediate phases during cycling. From a theoretical perspective, the formation energy (and voltage) are defined as internal energy differences between phases. Therefore, an accurate prediction of internal energy is crucial for the calculation of OCV. In this work, we present a theoretical framework that decomposes the internal energy of a given TMO into distinct contributions with clear physical significance. Specifically, we break down the energy into parameters that can be more easily calculated (compared to DFT) and obtained from experimental databases. From these parameters, we define a potential term that can be calculated for different compositions, and can be used for calculation of voltage profile.

\end{abstract}

\maketitle


\section{\label{sec:level1}Introduction}

Among the physical and chemical parameters that govern the performance of a battery cathode, regardless of the charge carrier (e.g., lithium or sodium), the open circuit voltage (OCV) is a fundamental factor. It directly influences the energy density of the cell and provides valuable insights into the cathode's chemistry and the oxidation states of its active species, making it a crucial parameter for evaluating and predicting performance\cite{urban2016computational, van2013understanding}. The OCV is inherently linked to the electronic structure of the cathode material. As such, modifications to the electronic structure through doping, substitution, or other strategies—can be effectively monitored and interpreted via changes in the OCV profile. Additionally, the presence of voltage plateaus in the OCV profile is of particular interest. These plateaus often correspond to phase transitions, redox reactions, or the formation of intermediate phases during (de)sodiation or (de)lithiation \cite{roscher2011ocv}. Understanding the origins of these features is important, as they offer critical insight into the underlying structural and electronic changes occurring within the cathode during cycling. From a thermodynamic perspective, the equilibrium cell voltage is defined by the difference in chemical potential, $\mu$, of the charge carriers between the cathode and the anode\cite{urban2016computational}. This relationship underscores the deep connection between voltage and material energetics.

\begin{equation}
V = \frac{\mu^{cathode}-\mu^{anode}}{z}
\end{equation}
where z is the transfered charge. The lithium chemical potential represents the change in the free energy of the electrode material with respect to lithium content. Integrating Equation (1) over a finite reaction range yields the average voltage as a function of the free energy change in the overall anode/cathode reaction, as described by the Nernst equation.
\begin{equation}
V = \frac{\Delta G_r}{z}
\end{equation}
Furthermore, under low-temperature conditions, the entropy term becomes negligible, leaving only the change in internal energy to determine the average voltage($\Delta G_r \approx \Delta H_r$). This approximation can be effectively used in Density Functional Theory in order to calculate the average voltage with equation:
\begin{equation}
V = \frac{E\left[\mathrm{A}_{x_1}[\mathrm{TM}]\mathrm{O}_2\right] - E\left[\mathrm{A}_{x_2}[\mathrm{TM}]\mathrm{O}_2\right] - (x_1 - x_2) E[\mathrm{A}]}{(x_1-x_2)}
\end{equation}
where the internal energies ($\mathrm{E[A_{x1}[TM]O_2],  E[A_{x2}[TM]O_2],  and,  E[A] }$) can be obtained via first principle method. and, $x_1 > x_2$. 
This implies that three DFT calculations can determine the average voltage of a given cathode. However, it has also been established that the voltage value is correlated with the electronic structure of cathode materials\cite{van2013understanding}. Layered transition metal oxides, in particular, present challenges in obtaining an accurate electronic structure description. As a result, voltage values often depend on the choice of exchange-correlation functional. An additional layer of complexity arises when attempting to reproduce the voltage profile. To achieve this, thermodynamically stable phases during charge carrier de-intercalation must be identified\cite{jain2011formation}. Consequently, the number of first-principles calculations required increases significantly to construct the convex hull of intermediate structures. Once the convex hull has been constructed, a piece-wise voltage profile can be obtained by evaluating the equilibrium voltage (Equation 3) between phases with adjacent charge carrier concentrations. 

In this work, we propose a computationally efficient semi-empirical model to predict the voltage plateau of cathode materials without the need for extensive first-principles calculations. Our approach relies on physically meaningful parameters such as ionization energies and electrostatic interaction energies, which can be derived from experimental data or low-cost computational methods. By focusing on the key energy contributions that govern phase stability and charge transfer during intercalation, the model enables rapid estimation of average voltages and voltage steps between intermediate phases. For the analysis, we use Transition metal (TM) oxides with the general chemical formula $\mathrm{A[TM]O_2}$ due to their popularity as a cathode material in recent battery materials research. A total of 14 compositions with 7 TM elements (namely Ti, V, Cr, Mn, Fe, Co and Ni) and two charge carriers (A = Na and Li) were chosen for the analysis.

\section{\label{sec:level2}Results and Discussion}

\subsection{\label{sec:level3}Potential Model}

The conventional approach of calculating formation energies and voltage profiles in battery electrode materials relies mainly on total energies obtained from DFT for various charge carrier (Li/Na) concentrations (Equation 3). However, it is important to recognize that the total internal energy output by DFT is not a singular, isolated physical quantity; rather, it is a combination of multiple contributions arising from the fundamental interactions within the system. These include the electrostatic (Coulombic) interactions among ions and electrons, the kinetic energy of the electrons, the exchange-correlation energy arising from quantum many-body effects, the ionization energies associated with electron transfer processes, and the lattice relaxation energy due to structural distortions accompanying changes in charge-carrier concentration. 
This presents a possibility of deconstructing the total energy of a system into different physically meaningful components. In the context of charge/discharge processes, where the system loses/gains charge carriers, we identify two components as being especially critical in governing the energy landscape: (i) the difference in electrostatic (Coulomb) energy between adjacent phases, which captures the redistribution of charge as Li/Na ions are extracted from the lattice; and (ii) the ionization energy, which reflects the intrinsic redox activity of the transition-metal ions (or other redox-active ions), determining the energetic cost or gain associated with electron removal.
 The electrostatic energy component can be calculated with high precision using the Ewald summation, which is well-established for periodic systems. Meanwhile, ionization energies are measurable quantities, either experimentally via techniques such as photoelectron spectroscopy or computationally through isolated atom or ion calculations. Crucially, when a charge carrier (Li/Na) is removed from the system, it is the ionization energy that largely dictates the internal energy change associated with this redox process. Therefore, by combining the Coulomb energy difference and the ionization energy between two adjacent lithiation/sodiation states, one can construct an approximate yet physics-based estimate of the total energy difference. This estimate can then serve as the basis for predicting formation energies and voltage profiles without performing expensive DFT total energy calculations for every intermediate state.
{\small
\begin{equation}
\Delta E_{\text{sys}} = E_{\text{col}}\left[\mathrm{A}_{x_1}[\mathrm{TM}]\mathrm{O}_2\right] - E_{\text{col}}\left[\mathrm{A}_{x_2}[\mathrm{TM}]\mathrm{O}_2\right] -  \mathrm{IE}[\mathrm{TM}]
\end{equation}
}
\begin{equation}
E_{\text{Inter}}\left[\mathrm{A}_{x}[\mathrm{TM}]\mathrm{O}_2\right] = E_{\text{col}}\left[\mathrm{A}_{x}[\mathrm{TM}]\mathrm{O}_2\right] + \left(\mathrm{IE}[\mathrm{TM}] \times n_I\right)
\end{equation}
\begin{figure}[htbp]
    \centering
    \includegraphics[width=\columnwidth]{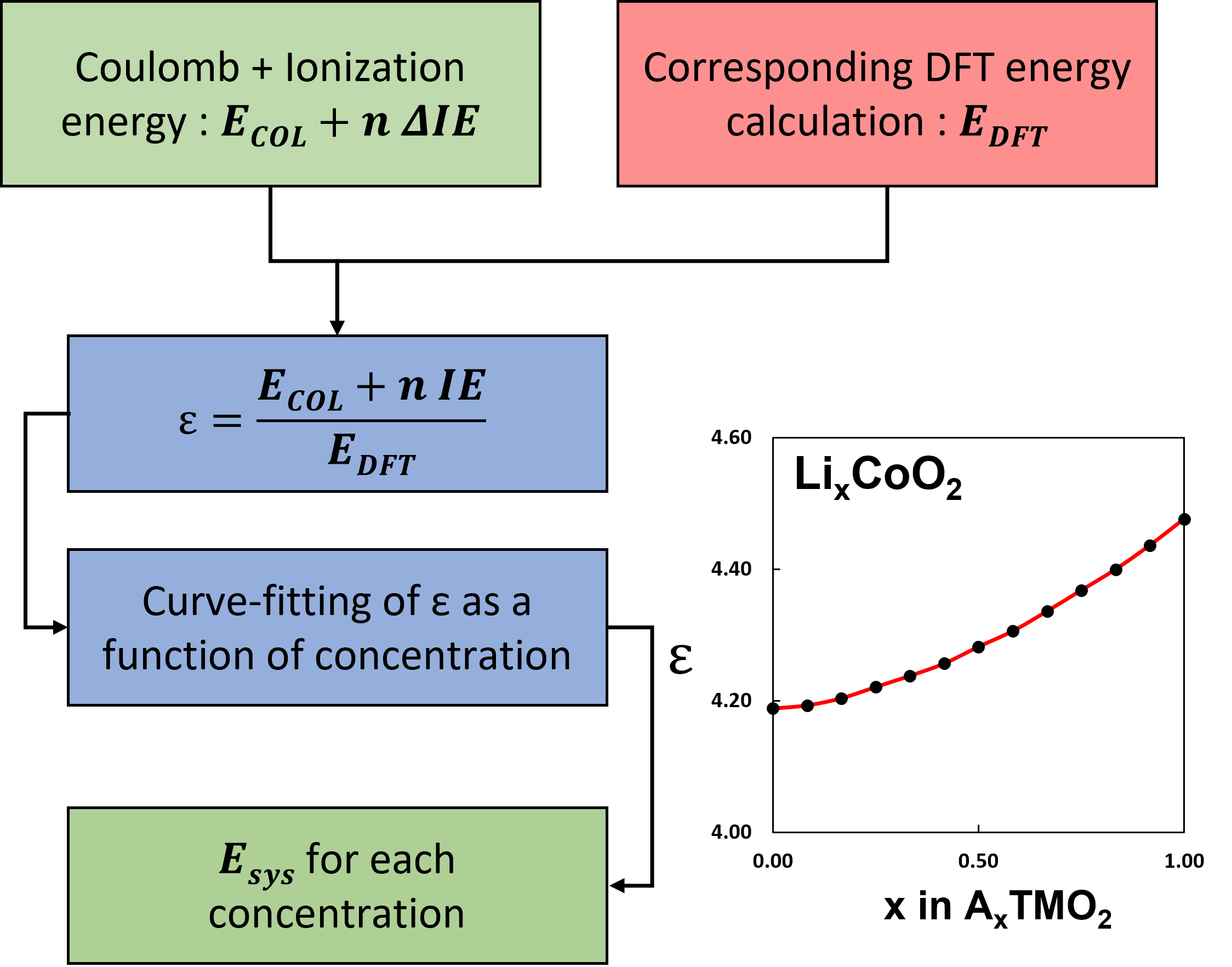}
    \caption{ Workflow for constructing $\epsilon$ vs Na/Li concentration plot, used for E$-{sys}$ prediction }
    \label{fig:voltage-ti1}
\end{figure}
\begin{figure}[b]
    \centering
    \includegraphics[width=\columnwidth]{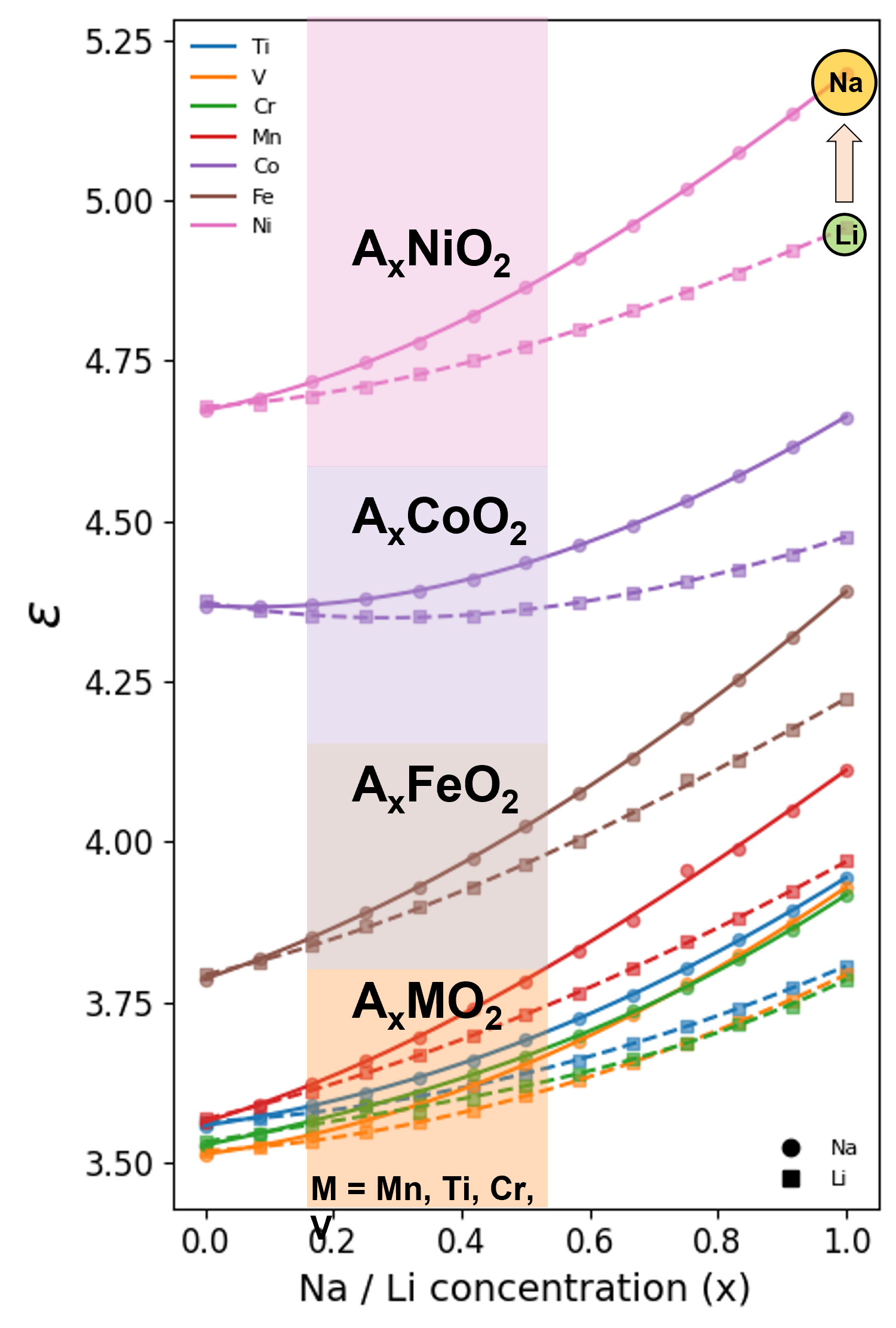}
    \caption{ Fitted $\epsilon$ of different Na/Li$_x$TMO$_2$ as a function of charge carrier concentration}
    \label{fig:voltage-ti2}
\end{figure}
where, $\Delta E_{sys}$ is the energy difference function between two adjacent phases, $E_{col}$ is the coulomb energy of a given phase, and $\Delta IE$ is the ionization energy correction. In order to impliment the ionization energy change more effectively, we adopted the following formulation. 
$E_{IE}$ is applied based on the number of charge carriers removed from the structure. As Na/Li is extracted, an equivalent number of TM ions are assumed to be oxidized from the 3+ to 4+ state to maintain charge neutrality. This redox energy contribution is estimated by multiplying the 3+/4+ ionization energy of the TM element by the fraction of TM ions oxidized per formula unit, which is proportional to the number of sodium vacancies. In this way, each partially desodiated state is assigned a physically motivated energy value that reflects both electrostatic and redox contributions. This is expressed in equation 5, Where $E_{inter}$ and $E_{col}$ are the approximate internal energy and Coulomb energy of a given phase $A_{x}[TM]O_2$, $IE[TM]$ is the ionization energy of redox-active TM ions (3+/4+ in this work) and, $n_I$ is the number of TM ions per formula unit, undergoing oxidation at the given charge carrier concentration. 


To validate the proposed energy decomposition framework, a diverse set of fourteen distinct transition metal oxide (TMO) compositions was selected for systematic analysis. Each compound adheres to the general stoichiometry $A_x[TM]O_2$, where $A$ denotes the charge carrier species either lithium (Li) or sodium (Na) and TM represents a transition metal chosen from the set {Ti, V, Cr, Mn, Fe, Co, Ni}. The variable $x$, indicating the charge carrier concentration, was systematically varied from 0 (fully delithiated/desodiated) to 1 (fully lithiated/sodiated). 
$E_{inter}$ from equation 5 was then calculated for all 14 compositions and their intermediate phases.
These computed energies ($E_{inter}$) were then fitted against reference DFT single-point total energies at the corresponding concentrations. For each composition, twelve intermediate charge carrier concentrations (evenly spaced between $x=0$ and $x=1$) were analyzed using this methodology. The residual differences between the approximated and DFT-computed total energies were captured via a correction term ($\epsilon$), which was used to enhance the accuracy of the energy profile.
\begin{equation}
 E_{I} [A_{x}[TM]O_2] = \frac{E_{col}[A_{x}[TM]O_2] + (IE[TM]\times n_I)}{\epsilon}
\end{equation}
\begin{figure}[htbp]
    \centering
    \includegraphics[width=\columnwidth]{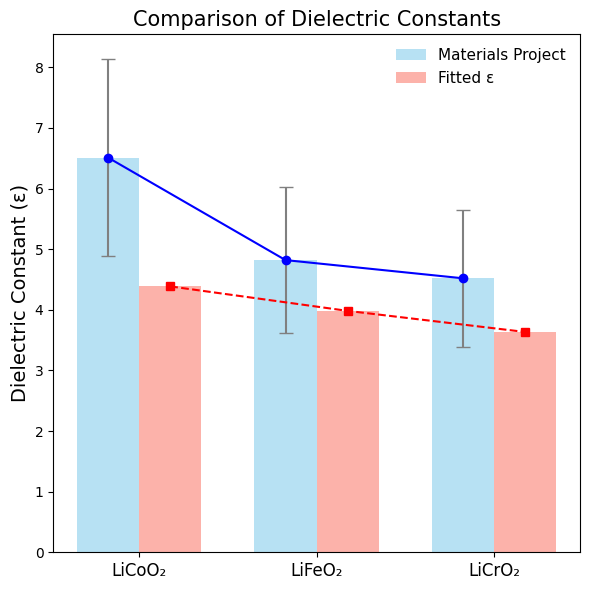}
    \caption{Comparson between $\epsilon$ values from materials project database and fitting parameter $\epsilon$ from thid work}
    \label{fig:voltage-ti4}
\end{figure}

Finally, $\epsilon$ obtained from this comparison was plotted as a continuous function of charge carrier concentration. This allows for interpolation of the corrected internal energy at any arbitrary concentration within the interval $0 \leq x \leq 1$, thereby enabling continuous estimation of formation energies and voltage profiles without requiring exhaustive DFT sampling. This continuous correction approach forms the basis of our semi-empirical (SE) model.
Figure 2 illustrates the variation of $\epsilon$ as a function of charge carrier concentration across all fourteen studied compositions. A consistent trend is observed wherein, for every concentration, the sodium-based ($Na_xTMO_2$) systems exhibit systematically higher values of $\epsilon$ compared to their lithium-based ($Li_xTMO_2$) counterparts. When looking at specific charge carrier concentrations, the difference in $\epsilon$ values between Na and Li systems becomes larger at higher concentrations. A comparison of the average $\epsilon$ values across different compositions reveals a broader trend: systems that contain transition metals with higher 3+/4+ ionization energies tend to exhibit larger average $\epsilon$ values.
\begin{figure*}[t!]
    \centering
    \includegraphics[width=\textwidth]{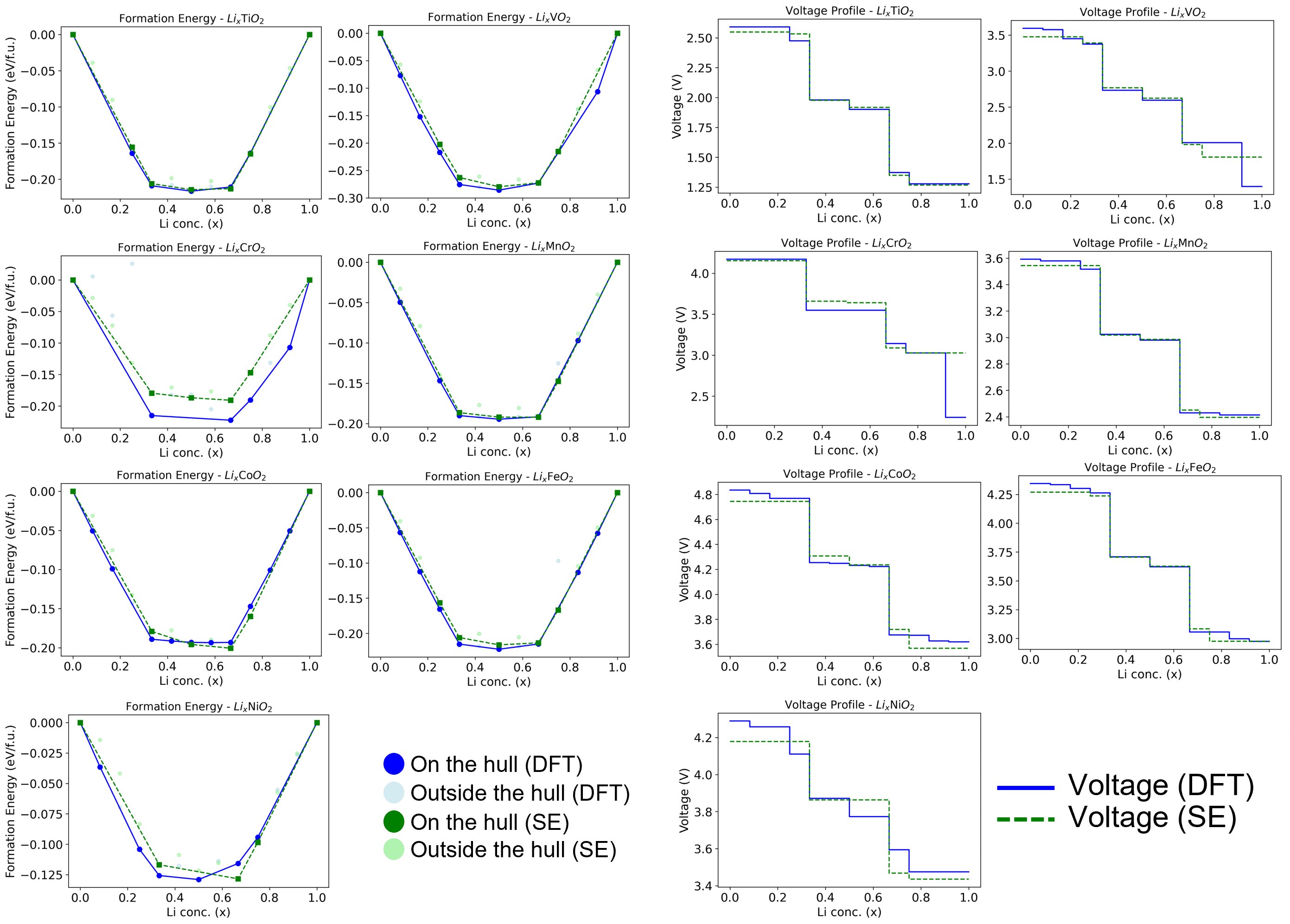}
    \caption{Formation energies and Voltage profiles of Li$_x$TMO$_2$ from SE and DFT calculations}
    \label{fig:voltage-ti5}
\end{figure*}
\subsection{\label{sec:level5} Physical significance of $\epsilon$}
In our semi-empirical (SE) model, the parameter $\epsilon$ was introduced as a concentration-dependent correction to the internal energy, aimed at capturing systematic differences between DFT-computed energies and their idealized, non-interacting ionic counterparts. Intriguingly, this parameter appears to possess a potential physical interpretation. Specifically, we propose that $\epsilon$ may correspond to the effective dielectric constant of the material system. This interpretation is grounded in the notion that the electrostatic interaction energy between charged species in a solid is screened by the material's dielectric response. In a classical Coulombic framework, the electrostatic potential energy scales inversely with the dielectric constant. Therefore, the ratio of the bare Coulombic (unscreened) interaction to the actual DFT total energy, which includes electronic screening effects, can be seen as a measure of the degree of dielectric screening in the material. 

\begin{figure*}[t!]
    \centering
    \includegraphics[width=\textwidth]{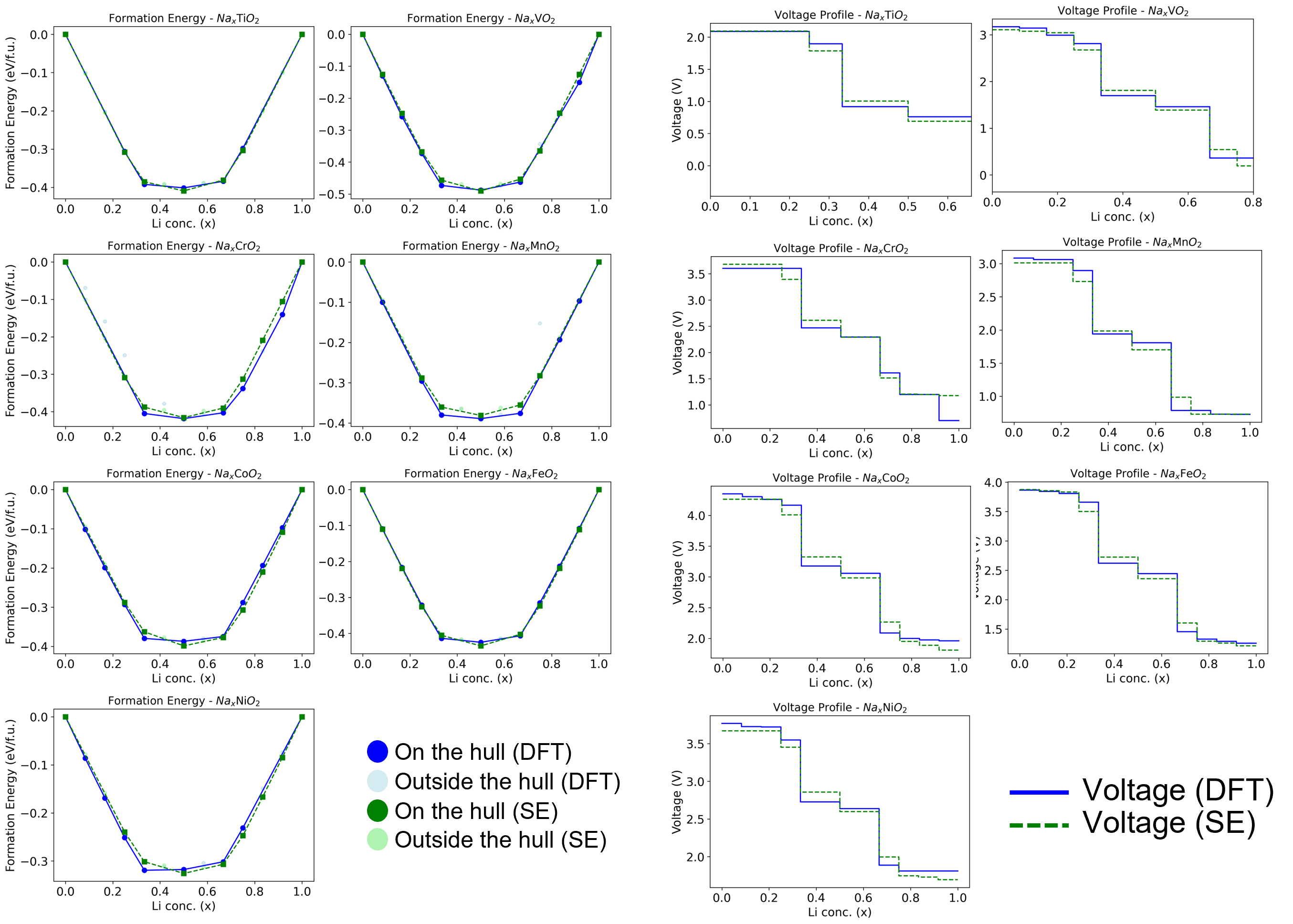}
    \caption{Formation energies and Voltage profiles of Na$_x$TMO$_2$ from SE and DFT calculations}
    \label{fig:voltage-ti6}
\end{figure*}

The fitted values of $\epsilon$ exhibit systematic and chemically intuitive trends across both Li- and Na-based transition metal oxides (TMOs), supporting its proposed physical interpretation. As shown in Figure 2, $\epsilon$ increases monotonically with increasing alkali content ($x$) in all systems studied. This trend is consistent with enhanced screening as more Li+ or Na+ ions enter the lattice, altering the local electronic environment.
A comparison between Na-based and Li-based compounds reveals that $\epsilon$ values are consistently higher for Na systems at the same level of intercalation and for the same transition metal species. For instance, at full intercalation ($x = 1$), Na–NiO$_2$ exhibits $\epsilon \approx 5.20$, compared to $\epsilon \approx 4.96$ for LiNiO$_2$. This difference arises from the larger ionic radius and greater polarizability of Na+, which leads to more effective dielectric screening in the Na-based hosts. Despite this difference in absolute values, the rate at which $\epsilon$ increases with $x$ (i.e., the slope of the $\epsilon$ vs. $x$ curve) remains remarkably similar between Na and Li systems. This suggests that while the screening strength differs, the underlying mechanism governing screening evolution is shared.
Transition metal identity also plays a key role. Systems based on Co and Ni show steep increases in $\epsilon$ with composition and reach the highest values overall. These elements are known for having accessible multiple oxidation states and relatively delocalized $d$-orbitals, which likely facilitate charge redistribution and electronic polarizability during intercalation. In contrast, Cr- and V-based systems show lower $\epsilon$ values and shallower slopes, indicating more rigid electronic environments and reduced screening capability. It is also noteworthy that the value of $\epsilon$ follows the electronegativity trend of the transition metals in the periodic table. This correlation arises because elements with lower electronegativity tend to form more polarizable bonds and less tightly bound electron densities, which in turn enhance the material’s ability to screen electric fields. 
To further test the hypothesis that $\epsilon$ captures electronic dielectric screening, we compared our fitted values with independently computed dielectric constants from the Materials Project database \cite{Jain2013}, which uses density functional perturbation theory (DFPT) to calculate static dielectric tensors. A direct comparison across all compositions was not feasible due to differences in space group symmetry among many database entries, which can significantly influence dielectric properties. However, we identified three Li-based layered oxides, LiCoO$_2$, LiFeO$_2$, and LiCrO$_2$, that share the same space group. This allowed for a controlled comparison across transition metals while minimizing structural effects. As shown in Figure 6, both our fitted $\epsilon$ values and the DFPT-calculated dielectric constants exhibit the same qualitative trend: Co-based compounds have the highest dielectric constant, followed by Fe and then Cr, which has the lowest. It is also worth noting that DFPT calculations tend to overestimate the electronic dielectric constant by approximately $25\%$. After accounting for this overestimation, the numerical values from the Materials Project fall within the same range as our model-derived $\epsilon$ values. This agreement supports the physical relevance of $\epsilon$ and validates its use as a proxy for dielectric screening.
Interestingly, since the refractive index $n$ is related to the dielectric constant by $n = \sqrt{\epsilon}$, the semi-empirical model provides a bridge between optical (electrodynamic) properties and electrochemical behavior. 
\begin{figure}[htbp]
    \centering
    \includegraphics[width=0.9\columnwidth]{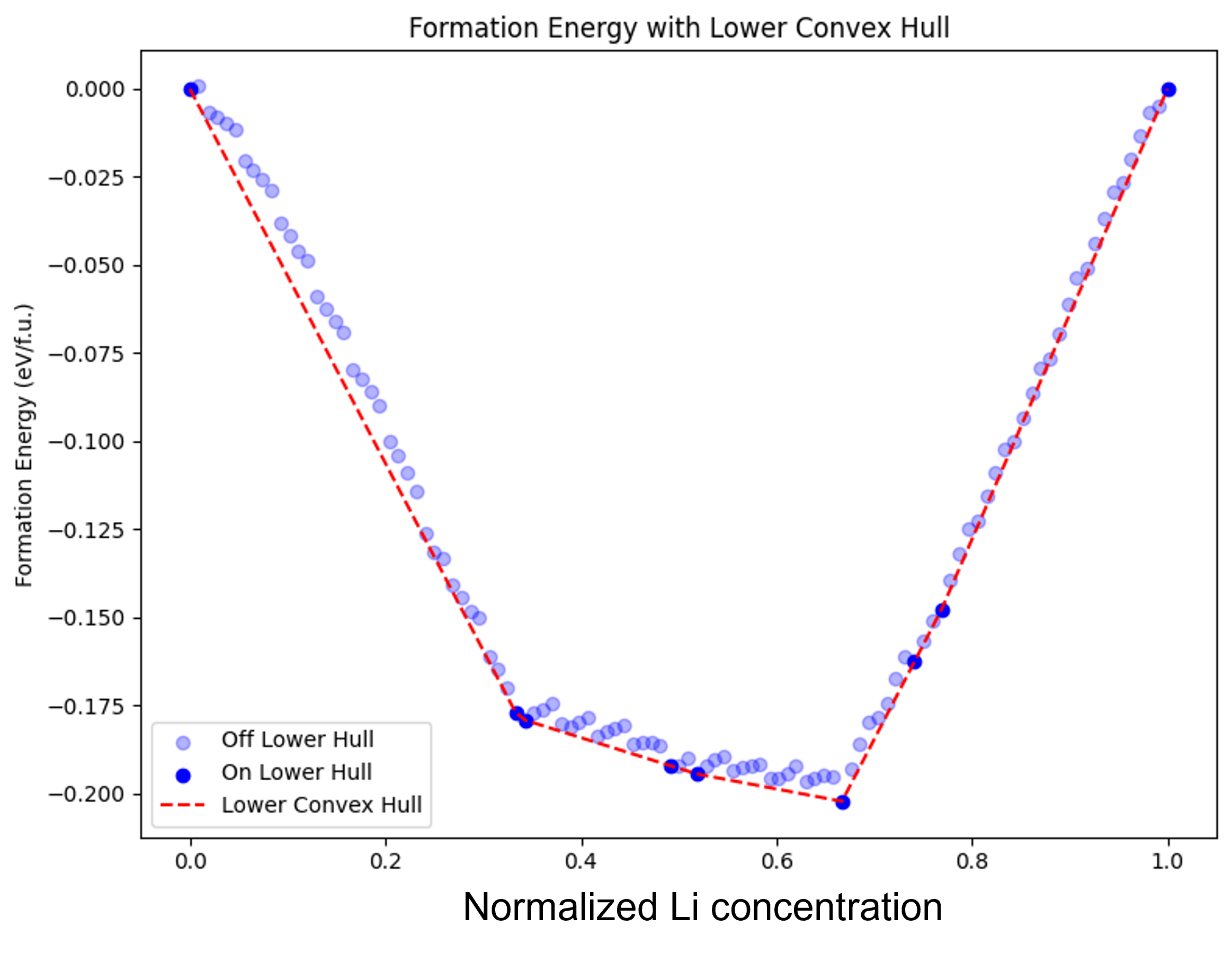}
    \caption{Convex hull of  Li$_x$CoO$_2$ with 110 intermediate phases}
    \label{fig:voltage-ti7}
\end{figure}
\subsection{\label{sec:level7}Convex Hull of the Formation Energies}

The formation energy of each compound within the Na–TM–O and Li–TM–O systems was derived from the total energy differences (E$_{sys}$ obtained using the semi-empirical (SE) model. These energies were used to construct compositional phase diagrams in the form of convex hulls for each of the 14 investigated compositions. 
The convex hulls in Figure 2 and 3 show that the SE model successfully reproduces the DFT hull for the Na and Li-based systems respectively.
To quantitatively assess the predictive capability of our semi-empirical (SE) model in estimating energy differences for arbitrary intermediate compositions, we applied the model to $Li_xCoO_2$ system. A total of 110 compositions with concentration range of $0 < x < 1$ were selected with fine resolution. For each composition, the fitting parameter $\epsilon$ was obtained by interpolating the continuous $\epsilon$ vs. Li concentration relationship derived from the fit shown in Figure 2. Using these interpolated $\epsilon$ values, the total energy difference ($\Delta E_\mathrm{sys}$) was calculated for each composition, incorporating the correction term into the base DFT energies. Subsequently, the formation energy for each composition was computed. This process enabled the construction of a high-resolution compositional phase diagram for the $Li_xCoO_2$ system. The calculated formation energies of the 110 intermediate states and the resulting convex hull are presented in Figure 5. 
The voltage profiles predicted by our semi-empirical (SE) model are fairly in quantitative agreement compared to those obtained from density functional theory (DFT) calculations for all studied Na- and Li-based transition metal oxide systems. The results of this comparison are presented in Figures 4 and 5 for charge carriers Li and Na respectively. The SE-derived voltage curves reproduce the general shape and average value of the DFT-calculated voltage profiles with reasonable accuracy. Both Na and Li based systems display close agreement in terms of the computed average voltages across the full concentration range. However, some deviations are observed in the detailed structure of the voltage profiles. These deviations are attributed to the simplified nature of the SE model, which, while incorporating concentration-dependent corrections via the fitted $\epsilon$ parameter, does not explicitly resolve electronic structure changes or localized redox events. In contrast, DFT calculations inherently account for these effects. Despite these limitations, the SE model demonstrates a high level of consistency in predicting average voltages, making it a computationally efficient alternative for large-scale screening and initial evaluation of voltage characteristics across complex compositional spaces. 

\section{\label{sec:level9}CONCLUSION}

We begin by considering a conceptual hypothesis for materials formation: imagine all constituent atoms in their neutral, free states, isolated and non-interacting placed together in a closed system. Upon confinement, these atoms undergo ionization, exchanging electrons to achieve stable electronic configurations and forming a solid compound. The resulting material is stabilized through long-range Coulombic interactions among the newly formed ions, alongside contributions from other energetic terms (both quantum mechanical and classical). Within the context of cathode materials, this simplified picture suggests that the predominant energetic contributions during phase transitions between adjacent compositions can be interpreted through changes in ionization energy and electrostatic interaction, both of which happens due to the ionization of TM and departure of the chrage carrier. Building on this idea, the semi-empirical (SE) formulation employed in this work was proposed to isolate these key contributions, allowing the estimation of formation energies and voltages by emphasizing ionic character and electrostatic stabilization across the composition space.
An essential component of the SE model is the fitting parameter $\epsilon$, which was introduced as a scaling factor to relate the computed electrostatic energy to the internal energy obtained from DFT.


One of the primary strengths of the SE model lies in its ability to compute energies for arbitrary intermediate compositions at negligible computational cost. This capability was demonstrated through the construction of a high-resolution convex hull for LCO with over 100 intermediate compositions. This makes the SE approach particularly attractive for large-scale screening and stability prediction in complex materials systems.Additionally, the model parameter $\epsilon$ shows strong resemblance to the effective dielectric constant, establishing a conceptual link between electrochemical behavior and optical/electrodynamic properties. Overall, the semi-empirical model offers a valuable balance between computational efficiency and physical accuracy. It provides a practical tool for rapid energy estimation and phase stability analysis in complex oxide systems. This framework provides a practical alternative for high-throughput screening of cathode candidates while maintaining a close connection to the underlying chemistry of transition metal oxides.

\section{\label{sec:level10}METHOD}

Conventional O3 type unit-cells were used for all $A_xTMO_2$ transition metal oxides (TMOs) where A=Li/Na and TM= Ti, V, Mn, Cr, Fe, Co and Ni. To find the DFT-based voltage profile and formation energies, single-point density functional theory (DFT) calculations were performed for 12 intermediate compositions of each 14 TMOs.  All first-principle calculations were performed using the generalized gradient approximation (GGA) described by the Perdew, Burke, and Ernzerhof (PBE)\cite{perdew1996generalized} with the Vienna Ab-initio Simulation package (VASP)\cite{kresse1993ab,kresse1996efficiency,kresse1996efficient}. The projector augmented wave method (PAW)\cite{blochl1994projector} implemented within VASP was used. A Monkhorst-Pack k-points (MPk) mesh of $2\times2\times1$, an energy cutoff of 520 eV as well as energy and force convergence criteria of $10^{-4} eV$ and 0.02 $eV$\AA$^{-1}$, respectively were used for all DFT calculation.
The electrostatic energy of all studied compositions were calculated by Ewald summation, implemented within GOAC code\cite{koster2025optimization}. The $\epsilon$ fitting was achieved by plotting the corresponding $\epsilon$ value as a function of 12 intermediate concentration of each TMOs. The $\epsilon$ value at each given concentration was calculated by the equation: 

\begin{equation}
\epsilon = \frac{E_{col}[A_{x}[TM]O_2] + (IE[TM]\times n_I)}{E_{DFT} [A_{x}[TM]O_2]}
\end{equation}
where, $E_{col}$ and $E_{DFT}$ are the coulomb energy and corresponding DFT single-point energy of a given phase, IE(TM) is the ionization energy of the transition metal and $n_I$ is the fraction of charge carrier that left the system to achieve the given charge carrier concentration. A second degree polynomial fit was then performed in order to interpolate the dataset, allowing rapid evaluation of total energy differences for any given adjacent phases.  \\

%

\end{document}